# Immutable Explainability: Fuzzy Logic and Blockchain for Verifiable Affective AI

Marcelo Fransoy[1], Alejandro Hossian[1], Hernán Merlino[1]
[1]Grupo GEMIS.BA – Universidad Tecnológica Nacional, Facultad Regional Buenos Aires (UTN FRBA), Buenos Aires, Argentina
afransoy@frba.utn.edu.ar*, ahossian@frn.utn.edu.ar, hmerlino@frba.utn.edu.ar

**Abstract** — Affective artificial intelligence has made substantial advances in recent years; yet two critical issues persist, particularly in sensitive applications. First, these systems frequently operate as 'black boxes', leaving their decision-making processes opaque. Second, audit logs often lack reliability, as the entity operating the system may alter them. In this work, we introduce the concept of Immutable Explainability, an architecture designed to address both challenges simultaneously. Our approach combines an interpretable inference engine—implemented through fuzzy logic to produce a transparent trace of each decision—with a cryptographic anchoring mechanism that records this trace on a blockchain, ensuring that it is tamper-evident and independently verifiable.

To validate the approach, we implemented a heuristic pipeline integrating lexical and prosodic analysis within an explicit Mamdani-type multimodal fusion engine. Each inference generates an auditable record that is subsequently anchored on a public blockchain (Sepolia Testnet). We evaluated the system using the Spanish MEACorpus 2023, employing both the original corpus transcriptions and those generated by Whisper. The results show that our fuzzy-fusion approach outperforms baseline methods (linear and unimodal fusion). Beyond these quantitative outcomes, our primary objective is to establish a foundation for affective AI systems that offer transparent explanations, trustworthy audit trails, and greater user control over personal data.

## 1. Introduction

In recent years, there has been a growing demand for virtual assistants capable of understanding and responding to human emotions. These systems are increasingly deployed in mental health, education, and customer support, and their adoption highlights both the potential and the limitations of affective technologies.

Currently, most affective AI systems remain, for the most part, "black boxes": it is often unclear how they arrive at their decisions. A second, perhaps more critical challenge concerns the lack of trust in audit logs. When a provider controls the models, the data, and the logging mechanisms, neither users nor regulators can verify that records are authentic or free from tampering. In practice, this results in closed data silos that hinder external auditing.

To address this dilemma, several authors have proposed alternatives. For example, Kushwaha (2025) suggests an immutable blockchain-based log in which every component—inputs, intermediate reasoning steps, and final outputs—is recorded. Their experiments show that this strategy improves traceability and accountability without incurring excessive costs. This line of inquiry aligns closely with the framework presented here.

To tackle the dual challenge of explaining AI decisions and ensuring those explanations remain unaltered, we designed an architecture built on two pillars. The first is a Fuzzy Logic inference engine, a "white-box" model interpretable by design that allows for the inspection of step-by-step reasoning. The second is a decentralized auditing subsystem leveraging blockchain technology and Self-Sovereign Identity (SSI) principles. This subsystem ensures that any trace generated by the system can be independently verified without relying on the provider's goodwill.

This integration leads to the central concept of our work: **Immutable Explainability**. Fundamentally, this combines a transparent, comprehensible inference process—leaving a clear record of decision provenance—with a cryptographic anchoring mechanism that protects the record against manipulation. Decoupling these responsibilities—where the AI provides the explanation and the blockchain ensures its verifiability—aims to establish systems that are genuinely auditable and trustworthy.

This research sits at the intersection of Explainable AI (XAI) and Distributed Ledger Technologies (DLT). Our objective is not merely to generate explanations, but to guarantee their integrity through cryptographic proof. Recent proposals, such as Parlak's (2025) BAXDT architecture, follow a similar trajectory by combining XAI and blockchain to record decisions on a public ledger. These approaches reinforce the necessity of advancing toward AI models whose explanations are not only clear but also verifiable and immutable.

Furthermore, the "heuristics-first validation" approach contributes significant methodological value. It applies a mature software design pattern to manage the complexity and risk inherent in secure multimodal AI systems, establishing an interpretable baseline before incorporating more computationally expensive neural models (Devlin et al., 2019; Fayek et al., 2022). Additionally, we utilize Whisper—the robust automatic speech recognition (ASR) model by OpenAI—as an instrumental ASR engine, without training or fine-tuning, to ensure the data pipeline remains reproducible and consistent (OpenAI, 2022; Hsu et al., 2021).

## 2. Related Work

2.1 Multimodal Emotion Recognition (MER). Currently, MER constitutes a pivotal area of research in machine-based emotion understanding. Existing approaches integrate multiple cues—acoustic, prosodic, and semantic—in an effort to minimize classification errors (Pei et al., 2024). However, a persistent challenge remains: noise. This issue is particularly acute when systems rely on text generated by ASR. Even when error rates are low, any distortion in transcription immediately compromises emotional coherence (Sahu et al., 2019).

To mitigate this vulnerability, several state-of-the-art techniques have been proposed. Many employ adaptive fusion, incorporating mechanisms that dynamically adjust the weight of each modality based on its momentary reliability. Prominent strategies include neural models that integrate auxiliary error-detection tasks and hierarchical attention architectures that redistribute importance across channels depending on their local trustworthiness (He et al., 2024).

Other strategies advance further by employing graph-based attention networks to model modality interactions, allowing contextual information to propagate between acoustic and textual nodes (Faiury et al., 2025). While these models achieve high performance, their weighting logic remains opaque: the system makes decisions but lacks granularity in explaining how they were reached. Consequently, they remain classified as "black-box" systems.

Our proposal is interpretable and auditable by design. Rather than delegating fusion to a complex statistical model, we employ a Mamdani-type fuzzy inference engine to determine the weight assigned to the textual modality (`w_text`) based on ASR confidence (`asr_confidence`). This enables a transparent implementation of the modality-gating principle: instead of optimizing weights automatically, we define explicit rules—e.g., "if ASR confidence is low, then the weight of the text modality decreases." The final decision emerges from a process that can be tracked step-by-step, audited, and understood without ambiguity.

In summary, our model prioritizes clarity and traceability over the sole maximization of performance metrics. While this may incur a slight trade-off in accuracy, it ensures transparency and paves the way for cryptographically verifiable explanations.

2.2 Explainable AI in Affective Computing. In affective computing, the necessity for explainability is increasingly critical. Decisions made by these systems can directly shape user emotional states and human-machine interaction (Hao & Liu, 2025). In the XAI literature, methods are typically categorized into two groups: post-hoc techniques applied after training, and models that are transparent ab initio.

Post-hoc methods such as LIME or SHAP are applied to pre-trained "black-box" models to explain specific decisions. While undeniably useful, these explanations are external approximations and do not fully reflect the model's internal logic.

By contrast, intrinsically interpretable or "white-box" models—such as decision trees, linear models, or rule-based systems—are transparent by design. Our work aligns with this category, employing a Mamdani fuzzy inference system (Zadeh, 1975). A key advantage of this approach is that each inference not only produces a numerical output but also leaves a clear trace of which rules were activated and to what degree, enabling step-by-step inspection of the reasoning process. This level of ante-hoc explainability is essential for debugging, validation, and fostering trust.

Beyond affective computing, other proposals aim to ensure AI traceability using blockchain technologies. Pegwar and Siddiqui (2025), for instance, propose recording decisions, inputs, and model artifacts on a blockchain to achieve comprehensive and transparent auditing. Although their approach is not emotion-focused, it reinforces the value of decoupling inference logic from verifiable traceability. This separation is precisely what we define as Immutable Explainability.

2.3 Privacy and Security in Voice Systems. Voice data is one of the richest—and most vulnerable—biometric modalities. Beyond semantic content, prosody and acoustic characteristics can reveal sensitive attributes regarding a speaker's health, emotional state, age, or geographic origin, presenting significant privacy risks.

These concerns stem not only from technical vulnerabilities but also from societal perceptions of AI. Zao-Sanders (2025) demonstrates that, even in everyday applications, users express growing distrust regarding how AI models handle their data and the transparency of their internal processes. This lack of clarity directly affects willingness to adopt voice-based technologies, where the involuntary exposure of sensitive attributes—such as emotion, health, or demographic traits—is particularly critical.

To mitigate these risks, the research community has explored various privacy-preserving techniques (Latif et al., 2023). Federated learning aims to train models without centralizing raw data, while differential privacy introduces statistical noise to protect individual identities. However, many of these architectures still rely on a central coordinator or fail to address the issue of verifiable auditing. Similarly, alternatives such as logs signed by a Certificate Authority (CA) or centralized Timestamping Authorities (TSA) continue to place full trust in the operating entity. If that entity is compromised or acts with malicious intent, audit integrity collapses. The need for a decentralized, immutable log has already been highlighted in IoT environments; for example, Kulothungan (2025) records AI inferences on a blockchain to provide a tamper-evident decision trace that multiple parties can verify without relying on a single authority.

Such vulnerability to alteration is untenable in high-stakes domains. In mental health, telemedicine, or forensic contexts, AI decision logs may be required as digital evidence. Accordingly, in the healthcare domain, blockchain has been combined with XAI to ensure trustworthy decisions: Bhardwaj et al. (2025) introduce a framework where each clinical explanation generated by XAI is anchored to a blockchain alongside interpretability metadata, allowing verification of the decision and its conditions, thereby strengthening clinical trust.

In such scenarios, it is a critical requirement that multiple stakeholders (e.g., the patient, the provider, and a regulatory body) can independently and reliably verify that an AI explanation has not been retroactively modified. This justifies the necessity of a distributed, immutable ledger over a centralized solution.

Our proposal aligns with an emerging paradigm that addresses this dependency: **SSI** built on blockchain infrastructure. Unlike traditional models, SSI empowers individuals with ownership and exclusive control over their digital identity and associated data (Zichichi et al., 2024). By grounding the architecture in Decentralized Identifiers (DIDs) and Verifiable Credentials (VCs), the system enables a new trust model. A user may receive a signed VC attesting that "their affective state was calm," where the proof of that trace resides in a public, immutable ledger independent of the service provider.

This approach represents a state-of-the-art alternative that prioritizes user sovereignty, third-party verifiability, and data minimization—principles essential for building genuinely secure and trustworthy voice systems—and justifies selecting this technology over centralized solutions.

### 3. Theoretical Framework
The methodological core integrates heuristic NLP—encompassing lemmatization, negation handling, intensifier processing, and both static and dynamic affective lexicons—**Mamdani-type fuzzy logic** for interpretability and graduality (Zadeh, 1975; Jang, 1993; Torres & Nieto, 2006), and late multimodal fusion weighted by ASR confidence and acoustic arousal (Van et al., 2025; Feng et al., 2024).

A Mamdani-type fuzzy inference system was selected as the backbone of the fusion process for two strategic reasons. First, fuzzy logic is inherently well-suited to handle the uncertainty and ambiguity characteristic of affective computing. Emotions are rarely absolute, discrete states; the ability of fuzzy logic to employ membership functions (e.g., "low arousal" or "medium confidence") captures this continuous variation explicitly (He, L., et al., 2024; Faiury, A., et al., 2025).

Second, unlike opaque neural models, a Mamdani engine provides interpretability by design. Each inference generates a readable audit trail—specifically, the linguistic rules that were activated and their corresponding firing strengths—as a natural byproduct of computation. This capability is essential for producing the explainable trace required by our concept of Immutable Explainability.

The heuristic framework further enables the refinement of rules and processing flows before incorporating computationally expensive neural models (Devlin et al., 2019; Fayek et al., 2022). Complementarily, this weighting mechanism serves as an interpretable implementation of the modality-gating principle, a technique explored in state-of-the-art neural architectures to mitigate the impact of noisy transcriptions (Rahman, W., et al., 2020).

Finally, system security relies on **PII (Personally Identifiable Information)** redaction and SHA-256 hashing for traceability (`txid`) and on-chain anchoring, with a planned integration of a SSI layer in a subsequent phase.

### 4. Heuristic Model Design
The architecture follows a Vertical Slice pattern composed of six subsystems: (i) Whisper ASR, (ii) audio-based emotion analysis, (iii) text-based emotion analysis, (iv) a fuzzy fusion engine, (v) an orchestrator/API, and (vi) an auditing and blockchain subsystem. The primary objective is to validate pipeline coherence and explainability using white-box modules and explicit rule sets, while deferring the integration of neural classifiers to a subsequent phase (Fayek et al., 2022; Van et al., 2025).

4.1 Design Principles. The architecture emphasizes interface-based decoupling (via EmotionResult), stable contracts, declarative configuration (YAML with local/environment overrides), automated testing (pytest), containerization (Docker/Colima), and observability (Prometheus). These principles facilitate the seamless substitution of backends with neural models without necessitating orchestrator refactoring or violating established contracts (Bohus & Horvitz, 2019).

4.2 Representation and Preprocessing.

Audio. Input audio is resampled to 16 kHz. The system extracts additional acoustic features: Root Mean Square (RMS) energy, Zero Crossing Rate (ZCR), and optionally Mel Frequency Cepstral Coefficients (MFCCs) if USE_MFCC=true and librosa is available. RMS and ZCR are heuristically combined to generate a preliminary arousal score, while MFCCs are used to derive a timbre_score that fine-tunes valence and arousal. The arousal signal is then smoothed using an Exponential Moving Average (EMA) with a configurable $\alpha$, stabilizing estimates across windows and turns. Additionally, Signal-to-Noise Ratio (SNR) estimation is incorporated as a metric of acoustic signal quality (Pan et al., 2024). SNR is computed heuristically over energy blocks using the lower (10th) percentile of the mean-squared energy, which enables estimation of background noise without additional models. This value, expressed in decibels, is stored in the audio-emotion metadata and used as an auxiliary factor to penalize ASR confidence (asr_confidence) under low-quality (<5 dB) or moderate (5–12 dB) conditions. This dynamic adjustment improves system robustness in noisy environments and reinforces the fuzzy-logic mechanism, as the weight of the textual channel decreases proportionally when estimated SNR is low—preventing decisions dominated by degraded transcriptions.

Text (Lemmatization, Negation, and Intensifiers). The text-based emotion module incorporates linguistic preprocessing to enhance robustness and alignment with affective lexicons. First, normalization and lemmatization are applied using the spaCy library (model es_core_news_sm) to reduce inflected forms to their base representation, facilitating polarity-lexicon lookups. Second, negation markers (e.g., no, nunca, sin) and their syntactic scope are detected to invert or attenuate the affective scores of lexical units within that scope. Third, intensifiers (e.g., muy, extremadamente, un poco) are identified to scale the magnitude of the affective score (heuristic multipliers > 1 for amplifiers and < 1 for attenuators). The textual output consists of: (i) a discrete probability vector per emotion (mirroring the audio ontology), (ii) continuous valence estimates normalized to [-1, 1], and (iii) metadata documenting lemmatized tokens, detected negations, and applied intensity multipliers. This refinement reduces false positives caused by polarity inversion and improves coherence during fuzzy fusion by providing textual features with greater fidelity to the underlying semantic content. To ensure reproducibility and transparency, the specific heuristics implemented in the audio and text backends are detailed below:

Audio — Extraction and Combination (Implemented Equations):

$rms\_norm = min(1, rms/(norm\_factor \cdot 0.92))$

$zcr\_raw = \#zero\text{-}crossing/\#samples$

$zcr\_norm = min(1.0, 10 \times zcr\_raw)$

$arousal\_combined = min(1, rms\_norm \cdot (0.9 + 0.1 \cdot zcr\_norm))$

$valence \leftarrow clamp(-1, 1, valence + (timbre\_score - 0.5) \cdot 0.2)$

$arousal\_combined \leftarrow min(1, arousal\_combined + timbre\_score \cdot 0.05)$

$arousal\_smoothed = \alpha \cdot current + (1 - \alpha) \cdot prev \text{ con } \alpha \text{ configurable.}$

Text — Intensifiers and Negations (Implemented Rules).

Intensifiers apply heuristic multipliers to the following token. Table 1 lists the intensifiers and their corresponding multipliers.

| Intensifiers | Multipliers |
|:---:|:---:|
| muy | 1.5 |
| extremadamente | 2.0 |
| sumamente | 1.8 |
| totalmente | 1.6 |
| algo | 0.8 |
| un poco | 0.7 |
| poco | 0.6 |

Table 1. Intensifiers and Multipliers

The weights and multipliers presented are heuristic and were calibrated for conservative behavior: arousal depends primarily on energy (RMS), and ZCR is introduced as a fine-grained adjustment factor (0.9/0.1) to capture variations in spectral excitation without making the system excessively sensitive to zero-crossing fluctuations. Lexical multipliers (intensifiers) were selected on a decreasing scale (2.0 → 0.6) to reflect typical linguistic strength patterns in Spanish. All parameters are configurable via YAML configuration files.

4.3 Whisper ASR (Instrumental). A small/base Whisper model is used to balance latency and quality. The system records `asr_confidence` for each segment. This confidence value acts as an input variable to the fuzzy engine, modulating the relative weight of the textual channel (Sahu et al., 2019; Hsu et al., 2021).

4.4 Continuous VAD (Valence, Arousal, Dominance) and Discrete Mapping. Each backend outputs continuous VAD values along with discrete probability distributions over {`alegría, tristeza, ira, miedo, asco, neutral`}. Valence is mapped to the range [-1, 1], and arousal/dominance to [0, 1]. The text modality contributes contextual valence, while the audio modality provides prosodic arousal.

4.5 Membership Functions and Universes. The variable `asr_confidence` ∈ [0, 1] with the following fuzzy sets: `LOW ~[0-0.5]`, `MED (0.3-0.8)`, `HIGH ≥0.65`. These are implemented using triangular and trapezoidal membership functions. Valence uses the categories `NEGATIVE (-1 to 0)`, `NEUTRAL (-0.25 to 0.25)`, `POSITIVE (0 to 1)`; arousal uses LOW/MED/HIGH over [0-1]. These ranges are empirically calibrated with controlled data and versioned in YAML for full traceability (Jang, 1993; Torres & Nieto, 2006).

4.6 Rule Base and Operators. IF–THEN linguistic rules with min t-norm and max s-norm; max aggregation and centroid defuzzification. Examples: (R1) `IF asr_confidence is HIGH AND valence is POSITIVE → w_text HIGH`; (R2) `IF asr_confidence is LOW AND arousal is HIGH → w_text LOW`; (R3) `IF asr_confidence is MED AND arousal is MED → w_text MED`; (R4) `IF valence is NEGATIVE AND arousal is HIGH → w_text LOW`. Conflicts are resolved via activation and centroid selection, and the set of activated rules is stored for each event (Hao & Liu, 2025).

4.7 Probabilistic Fusion and VAD. Fusion between the text and audio channels is performed as a convex combination of their discrete emotion distributions:
$Probs_{final} = w_{text} \cdot Probs_{text} + (1 - w_{text}) \cdot Probs_{audio}$

where $w_{text} \in [0, 1]$ is determined by a Mamdani-type fuzzy inference engine that takes as inputs ASR confidence (`asr_conf`), arousal, and valence.

Complementarily, a heuristic Multimodal Coherence Index ($C$) was implemented to quantify the consistency between the audio and text channels. The index is computed from the absolute differences in valence and arousal across modalities, normalized to their respective ranges:

$$C = 1 - \frac{(\frac{|valence_{audio} - valence_{text}|}{2} + |arousal_{audio} - arousal_{text}|)}{2}$$

such that $C \in [0, 1]$. High values ($C \geq 0.7$) indicate emotional coherence, whereas low values ($C \leq 0.4$) suggest misalignment or ambiguity. This index is recorded in the audit logs and monitoring metrics (gauge `cross_modal_coherence`), enabling analysis of correspondence between prosodic tone and semantic content. It can also serve as a control variable in the fuzzy weighting mechanism, reducing textual weight when modalities exhibit low coherence.

For VAD, the continuous output of the audio channel is used as the primary arousal signal (clamped to [0,1]), while the final valence value is computed as the mean of the estimates from both channels (mapping [0,1]→[−1,1] when needed). Instead of binary "hard gating," the system applies soft gating through fuzzy rules: when `asr_conf` is low, the rules tend to reduce $w_{text}$ and increase the audio contribution; when `asr_conf` is high, the engine favors the text. A degradation/fallback mechanism is implemented: if the fuzzy engine fails, a linear fusion based on `asr_conf` is used, and the audit log records the detected mode and applied weights to support observability.

4.8 Explainability and Auditing. Each inference preserves a complete trace for auditing and explainability. The fuzzy engine returns, in addition to the numerical value $w_{text}$, the input membership degrees (μ for each label), the fired rules with their activation strengths, and the clipped output sets; this information is stored in `stored_event["fusion_fuzzy"]`. This level of detail endows the audit record with semantic comprehensibility. Together, these elements constitute what we term Immutable Explainability: the fuzzy logic engine generates an interpretable audit trail, and the cryptographic subsystem seals it with a `txid` that guarantees immutability. This combination of interpretability and immutability forms the foundation of a genuinely auditable and trustworthy system. To support both human inspection and automated auditing, the system exports an explainability artifact for each inference, containing the evaluated rules and their activation degrees (`fired_rules`). This artifact is stored in JSON format and, when possible, accompanied by a heatmap (PNG image) representing a rules–conditions matrix colored by firing strength (0..1).

The JSON record includes: the `txid`, the list of `fired_rules` (with if, then, and strength fields), the engine inputs (`asr_conf, arousal, valence`), and the out_sets. The heatmap provides a fast visual cue of which conditions most strongly influenced the resulting `w_text`. These artifacts are stored in the audit repository (`/audit/fired_rules/`) and versioned alongside each event using SHA-256. The generated figures enable both qualitative and quantitative analysis (e.g., grouping the most frequently activated rules by cohort) and are valuable for justifying system decisions to regulators or ethics committees.

The following is an example of the `fusion_fuzzy` output:

- `inputs`: displays the raw signals (`asr_conf`, `arousal`, `valence`).
- `fired_rules`: lists the evaluated rules with their firing degree (`strength` = μ min across the rule's conditions).
- `out_sets`: shows the maximal aggregation per output set (`low/mid/high`) prior to defuzzification; the final `w_text` is the centroid resulting from defuzzification.

```
{
  "w_text": 0.5843812629945782,
  "details": {
    "inputs": {
      "asr_conf": 0.9582073547338185,
      "arousal": 0.12,
      "valence": 0.02
    },
    "fired_rules": [
      { "if": ["asr_conf is high"],          "then": "w_text is high", "strength": 0.9162 },
      { "if": ["arousal is low", "valence is pos"], "then": "w_text is high", "strength": 0.0 },
      { "if": ["valence is neu"],            "then": "w_text is mid",  "strength": 1.0 }
    ],
    "out_sets": {
      "low": 0.0,
      "mid": 1.0,
      "high": 0.9164147094658042
    }
  }
}
```

Example of Fuzzy Engine Output

The orchestrator redacts PII before persisting the event as JSONL and computes the corresponding txid (the SHA-256 hash of the redacted JSON). The stored fields include: `asr_conf`, `emotion_audio_conf`, `emotion_text_conf`, weights (`w_text/w_audio`), `fusion_fuzzy` (*inputs/fired_rules/out_sets*) and, when applicable, acoustic metadata (`arousal_raw`, `zcr_raw`, `zcr_norm`, `timbre_score`, `mfcc_present`, `arousal_smoothed`).

4.9 Orchestration and Fault Tolerance. The orchestrator controls the execution of components and exposes latency metrics. In this phase, fault tolerance relies on: (i) latency and error instrumentation, (ii) centralized exception handling, and (iii) a linear-fusion fallback mechanism in the event of fuzzy-engine failure. Circuit-breakers, automatic retries, and per-component timeouts are not implemented; nor is VAD segmentation performed at the orchestrator level (input is processed as a full file/turn).

4.10 Observability. The system provides Prometheus metrics, including latencies (ASR, audio emotion, text emotion, fusion), counters (`pii_redactions_total`, `pipeline_errors_total`), and gauges (`audio_snr_db`, `cross_modal_coherence`). All metrics incorporate `model_size` and `run_id` labels—sourced from the centralized configuration—facilitating cohort analysis (e.g., by noise level, accent, or model version). Latencies are instrumented using wrappers (`timeit`).

The `audio_snr_db` gauge records the estimated SNR in decibels for each execution, enabling correlations between signal quality, ASR performance, and fuzzy-engine behavior. The `cross_modal_coherence` gauge monitors affective alignment between the audio and text modalities, supporting multimodal consistency studies. JSON logs include structured fields (vad, metadata, probs), which—together with the JSONL audit trail—enable reconstruction of the explanation for each decision and its acoustic context.

4.11 Empathetic Response and Guardrails. User responses are generated via templates conditioned on the dominant emotion (`_plan_response`), applying simple attenuators to avoid unnecessary escalatory replies. Escalation routes and conceptual thresholds for risk states (e.g., expressions of health risks or self-harm) have been defined, although the automatic activation of human operators and their operational integration form part of the next-stage roadmap.

Guardrails execute immediately after probability fusion and before response generation. Simple rules based on probabilities (e.g., fear > 0.7) and sensitive-keyword detection within the text are evaluated. If a risk is detected, an escalation block is appended to the audit event (`stored_event`), triggering a notification to an operator or a message to a topic (via webhook if `escalation_webhook` is configured). All escalation decisions are recorded in the audit log (Lawrence et al., 2024). Figure 1 illustrates the complete system architecture.

4.12 Limitations of the Heuristic Phase. The current design is a heuristic Minimal Viable Product; consequently, it does not capture complex semantic nuances (e.g., sarcasm, irony), and its ability to distinguish between positive arousal (joy) and negative arousal (anger) is constrained by the nature of the acoustic signals used. The use of MFCCs and ZCR provides modest improvements to the acoustic signal, but accuracy will remain limited until the backends are replaced with neural models—BETO/RoBERTa-es (Liu et al., 2019) for text, and Wav2Vec2/WavLM/HuBERT for audio—while preserving the same API and the established EmotionResult and auditing interfaces (Devlin et al., 2019; Hsu et al., 2021; Fayek et al., 2022).

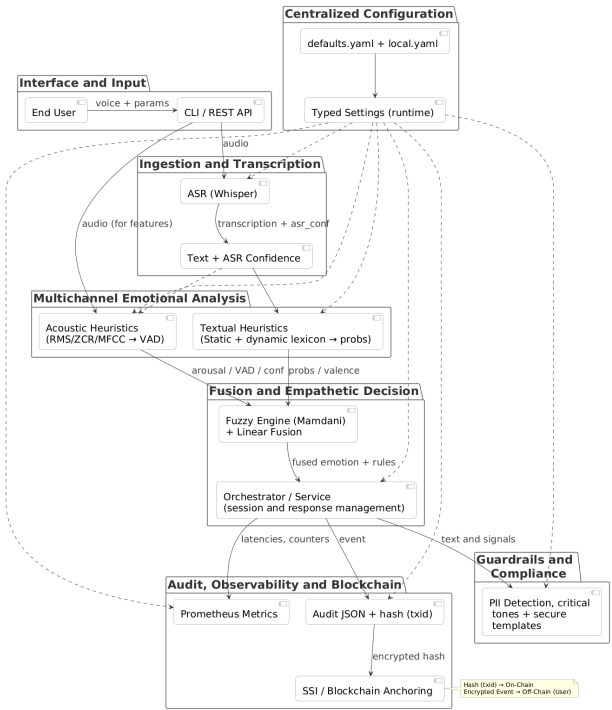

Figure 1 — Architecture of the proposed framework

4.13 Decentralized Auditing Layer. This layer serves as the architectural pillar that resolves the issues of trust and verifiable auditing (the "data silo"), effectively realizing the concept of Immutable Explainability on the blockchain. This technique was selected for two fundamental reasons:

Efficiency and Cost: Storing small hashes (32 bytes) on-chain is far more economical and faster than storing full audit JSON records.

Privacy: Sensitive data (the audit trace, even in redacted form) remains off-chain; only an immutable proof of existence is stored on-chain. The complete trace (the human-readable JSON) remains encrypted off-chain and under the user's sovereign control.

The mechanism operates as follows: each pipeline execution generates a redacted event (`stored_event`) and computes its cryptographic hash (`txid = SHA-256(...)`). This `txid` is anchored in a smart contract deployed on a public blockchain (Sepolia Testnet).

A key aspect of this design is that the complete JSON (the readable audit trace) remains encrypted off-chain and under the user's sovereign control—for example, in the InterPlanetary File System (IPFS) or a personal data vault. The public chain stores only the immutable proof of existence. It is important to note that although this scheme establishes the cryptographic foundations for data sovereignty, the present work is limited to integrity anchoring and does not implement a full SSI ecosystem. The management of DIDs and VCs lies outside the scope of this work and is reserved for future stages of system evolution.

It is crucial to recognize the limitation of this approach: the blockchain guarantees only the integrity and the proof of existence of the explanation; it does not store the explanation itself. Verification requires the off-chain holder of the data (e.g., the user) to voluntarily present it so that its hash can be compared to the on-chain record. Although emerging alternatives for verifying AI computations—such as zk-SNARKs or ZK attestations—may eventually enable proof of model execution without revealing underlying data, our hash-anchoring strategy represents a pragmatic and robust solution for this work's specific objective: ensuring the integrity of an already-generated explainable trace.

This integration constitutes the operational realization of our proposal: we decouple system operation from the verification of its audit trail. A dedicated module manages the Web3 connection, signs the transaction, and returns the anchoring status (disabled, submitted, anchored), allowing the user (or an authorized third party) to cryptographically verify that the audit trace (`fusion_details`) has not been altered—without relying on the provider's honesty.

## 5. Evaluation and Preliminary Results

5.1 Experimental Design. End-to-end coherence and rule stability under controlled perturbations were verified. The test suite consisted of the Spanish MEACorpus 2023 (Pan et al., 2024), comprising 5,129 audio samples; processing was performed in single-sample batches on an 8-thread CPU using FP32 precision. The observed variables included: per-component latency, stability of `w_text` under noise/transcription degradation, and consistency of activated rules.In addition to empirical evaluation, unit tests (pytest) were implemented to cover the heuristic modules (text, audio, fuzzy fusion, guardrails, PII redaction). This ensures automatic regression testing as neural backends are introduced and documents the behavioral contract of each component.

5.2 Metrics. Across repeated runs, average latencies were: ASR ≈ 3.9 s, fusion < 0.05 s, and emotion analysis < 0.01 s, resulting in a total end-to-end latency of ≈4.0s per request. Metrics were exposed through Prometheus with buckets showing bounded ASR queues and low variance in the fusion stage.

5.3 Ablations. (A1) No text channel: resulted in an increase in false positives for high-valence emotions. (A2) No audio channel: revealed sensitivity to high Word Error Rate (WER). (A3) No gating via `asr_confidence`: led to unstable decisions. (A4) No fuzzy engine (fixed weight): caused a loss of adaptability. Overall, the fuzzy engine demonstrated the best trade-off between stability and sensitivity to reliable signals (Feng et al., 2024; Sahu et al., 2019).

5.4 Comparative Quantitative Evaluation. To quantify the improvement introduced by the fuzzy fusion engine relative to simple baselines, an evaluation was performed on the Spanish MEACorpus 2023. Three baselines were considered:

(i) Text Only: text-based emotion module fed with Whisper's noisy transcription (real ASR).

(ii) Audio Only: audio-based emotion module.

(iii) Simple Linear Fusion: a convex mixture defined $\left(P_{final} = asr\_confidence * P_{texto} + (1 - asr\_confidence) * P_{audio}\right)$

Results are summarized in Table 2. Precision, recall, and F1-score are reported using both macro and weighted averages.

| Model | Accuracy (macro) | Recall (macro) | F1 (macro) | Accuracy (weighted) | Recall (weighted) | F1 (weighted) |
|---|---|---|---|---|---|---|
| Text Only | 0.456 | **0.437** | 0.388 | 0.454 | 0.424 | 0.396 |
| Audio Only | 0.082 | 0.155 | 0.107 | 0.167 | 0.297 | 0.213 |
| Linear Fusion | 0.451 | 0.426 | 0.388 | 0.446 | 0.425 | 0.408 |
| Fuzzy Fusion | **0.456** | 0.422 | **0.394** | **0.455** | **0.438** | **0.429** |

Table 2. Measurements of the 3 baselines and the proposed model

Figure 2 visualizes the weighted F1 comparison for the four evaluated approaches, while Figures 3–6 present the class-normalized confusion matrices.

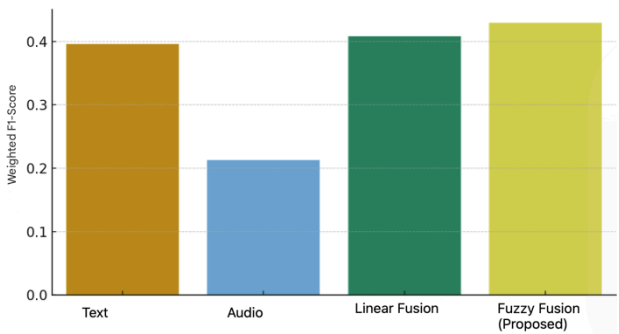

Figure 2. Comparison of F1-Score Weighted by Model

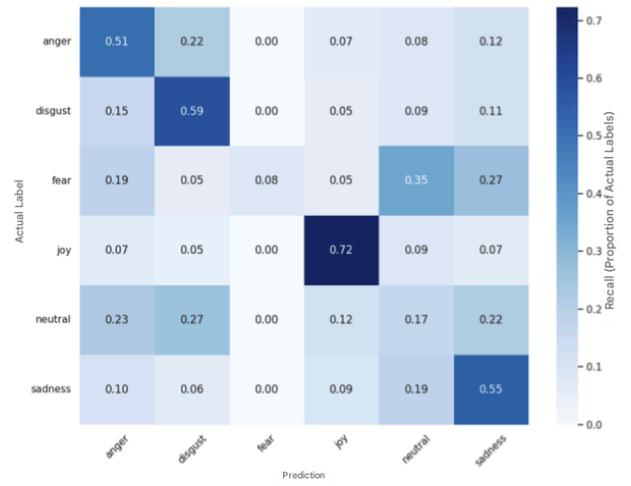

Figure 3. Confusion Matrix - Text Only

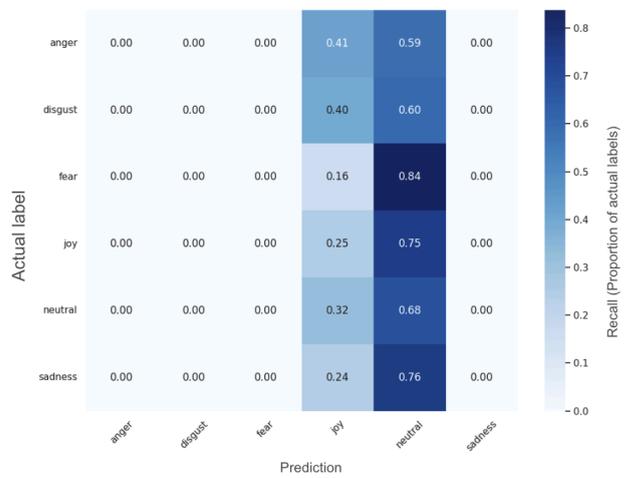

Figure 4. Confusion Matrix - Audio Only

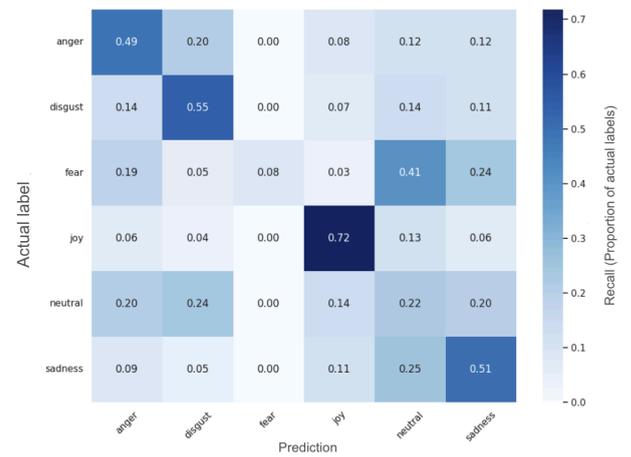

Figure 5. Confusion Matrix - Linear Merge

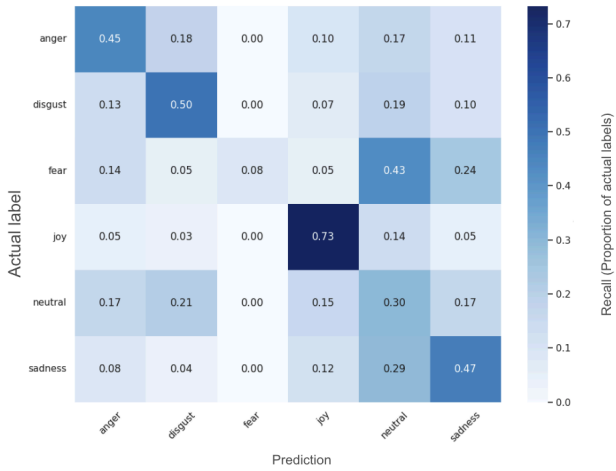

Figure 6. Confusion Matrix - Fuzzy Fusion (proposed)

Although the fuzzy fusion method continues to outperform the baselines in weighted F1, macro F1, and weighted recall, the text-only model retains a slight advantage in macro recall and weighted precision. This occurs because the text modality—even when fed noisy transcriptions—maintains a strong bias toward the majority classes (disgust and neutral), whereas the fusion mechanism distributes probability across anger, sadness, joy, and neutral, accepting a few additional false positives. We prefer this balance, as prioritizing the detection of critical affective states is more valuable than maximizing weighted precision.

The confusion matrices also show that the fuzzy engine reduces systematic errors between closely related classes (e.g., neutral vs. disgust) and maintains stability when ASR confidence decreases.

This quantitative evidence reinforces the "heuristics-first validation" philosophy: modeling uncertainty through explicit linguistic rules provides robustness without sacrificing interpretability or traceability.

5.5 Robustness. When `asr_confidence` < 0.6 and estimated arousal is high, the fuzzy engine decreases `w_text` and prioritizes the acoustic modality, preventing noisy text from dominating the decision. Rule activation remained stable across executions, improving explainability (Hao & Liu, 2025)

5.6 Case Study Analysis. To illustrate how the fuzzy engine behaves under real conditions of conflict and ambiguity, two representative examples were selected from the full set of 5,129 samples. These cases demonstrate how the system uses ASR confidence and fuzzy rules to dynamically weight the text and audio modalities. Table 3 summarizes the analyzed cases.

Case 1 — Correction Over Baselines. All baselines fail: the text module predicts anger, the audio module predicts neutral, and linear fusion inherits the text error. The fuzzy engine, however, produces joy, which is correct. Although ASR confidence is high (0.85), the `fusion_details` reveal rules that reduce `w_text` (≈ 0.63) because valence lies near the neutral zone. By leveraging the full probability vectors (`prob_text` and `prob_audio`), the fuzzy mixture recovers the secondary probability mass for joy and corrects the error. This clearly illustrates how the pipeline can recover correct predictions even when both individual modalities fail.

Case 2 — Conflict Resolution: Text is correct (sadness), while audio is incorrect (joy). With `asr_confidence` = 0.85, the rules prioritize text and set `w_text` ≈ 0.62. The fuzzy engine suppresses the incorrect audio hypothesis, and both linear and fuzzy fusion give sadness. This validates the core design principle: using ASR confidence as an arbiter to determine which modality should dominate, protecting the system from misleading prosody contradicting a reliable transcription.

Conclusion. The two cases reveal the dual behavior of the fuzzy engine with real ASR: it (1) corrects errors when all baselines fail (Case 1), and (2) resolves conflicts by prioritizing the more reliable modality (Case 2). All details are recorded in `fusion_details` (inputs, activated rules, weights) and can be audited via the on-chain `txid`.

5.7 Evidence of On-Chain Anchoring. Beyond quantitative metrics, we instrumented a verifiable auditing flow: for short runs, HashAnchored transactions are issued to record the `txid` of the `fusion_details`. The script `verify-txid-anchorage.py` enables any third party to confirm that (i) the local JSON has not been altered (matching hash), and (ii) the entry was anchored on Sepolia (`block_number`, `tx_hash`, `sender`).

Latency and Gas Analysis: The impact of the blockchain component on real-time flow and its economic feasibility was evaluated.

(i) Latency (Asynchronous Decoupling). The blockchain anchoring is designed as an asynchronous background task. While the user's interaction cycle (ASR + Fusion + Response) has an end-to-end latency of ≈ 4.0 s, the transaction submission occurs in a parallel thread after the response is delivered. Thus, the added user-perceived latency is essentially zero (0 ms). Block confirmation time on Sepolia/Ethereum (≈ 12 s) affects only the public availability of the audit record, not interaction fluency.

| Case | id (file) | ground truth | pred_text | pred_audio | pred_linear | pred_fuzzy | Input Data (Fuzzy) | | |
|---|---|---|---|---|---|---|---|---|---|
| | | | | | | | asr_confidence | audio arousal | audio valence |
| #1 Fuzzy Enhancement (correction) | d039c0 a8-d37 6cd8d. mp3 | joy | anger (failure) | neutral (failure) | anger (failure) | joy (success) | 0.85 (high) | 0.636 (medium-high) | 0.0 (neutral) |
| #2 Multimodal Conflict (text priority) | a7b075 71-e63 8f157. mp3 | sadness | sadness (success) | joy (failure) | sadness (success) | sadness (success) | 0.85 (high) | 0.919 (very high) | 0.0 (neutral) |

Table 3. Summary of cases analyzed

(ii) Gas Fees (Cost Metrics). The hash-write operation (`store_txid`) in the smart contract consumes ~47,000 gas units per event. Testnet (Sepolia): no monetary cost; ideal for development. Mainnet Projection (Layer 1): assuming a 50 gwei gas price and ETH at USD 3,445, the per-anchor cost is ≈ USD 8.08 (47k × 50 gwei). Scalability Strategy: For large-scale deployment, the design considers migrating to Layer-2 networks (Polygon, Arbitrum) or using Merkle aggregation, where hundreds of events are grouped into a single Merkle root before anchoring. This reduces marginal cost to negligible levels (< USD 0.01 per event), ensuring long-term viability.

This step completes the cycle of Immutable Explainability:

fuzzy engine → auditable trace → hash anchored on blockchain.

5.8 Threats to Validity and Scope of the Heuristic Model. It is essential to contextualize this evaluation within project objectives. The fuzzy-fusion model evaluated here is inherently heuristic. Its primary purpose is not to establish a new absolute performance benchmark but to validate the fusion architecture (the fuzzy rule engine) as a viable arbitration mechanism before moving to more complex implementations.

Key threats to validity include:

Heuristic Model Scope. The current approach is consistent with this early phase but is not representative of the final deployment. Future work will replace modular components (text and audio classifiers) with end-to-end neural models operating on the same Whisper ASR output. These models are expected to generalize better, compensate for ASR noise, and surpass the heuristic benchmark.

Generalization Validity (Cross-Corpus). Although the full MEACorpus was used, performance has not been evaluated on other Spanish speech corpora.

External Validity (End-Users). Results are based on a static dataset; evaluation with real users in interactive settings remains pending.

## 6. Discussion

Evaluation on the full MEACorpus under realistic conditions — Whisper transcriptions with variable `asr_confidence` — shows mixed signals. The fuzzy pipeline achieves a recall of 43.8% (weighted F1 = 0.429) but still leaves a "consistent error" of 33.3% (1,708 samples where all variants, including the proposed system, fail), indicating the high multimodal ambiguity of the corpus.

However, the objective of this phase was not to solve 100% of the samples but to verify that fuzzy logic arbitrates better than trivial fusion strategies. In this respect, the data are conclusive:

Fuzzy arbitration adds real value. Among the 5,129 cases, the pipeline corrects:

• 168 errors made by linear fusion,

• 287 errors missed by the text model,

• and 162 instances where both text and linear fusion fail simultaneously.

Correction behavior persists. Case 1 shows that, even with ASR noise, the fuzzy engine can recover joy by leveraging full probability vectors rather than the top-1 label.

Conflict handling holds. Case 2 confirms that ASR confidence governs modality weighting: when transcription is reliable but prosody misleading, the fuzzy engine suppresses the incorrect audio hypothesis.

This strongly supports the heuristics-first approach: even without neural models, the fuzzy engine adds resilience and preserves explainability, establishing a reproducible benchmark for future work.

The most important finding, however, is the system's auditing capability. Unlike an end-to-end neural "black-box" model, each `fusion_details` record provides a readable trace specifying which rules fired, with what strength, and why the final weights were assigned.

This property is the central pillar of the proposed architecture. By generating an auditable justification for each affective decision—and anchoring it on blockchain—the explanation becomes Immutable. This proof-of-concept thus serves as the foundational validation of this work's core idea: Immutable Explainability.

### 7. Conclusions and Future Work

7.1 Conclusions. This work demonstrates that the hybrid affective architecture is effective in practice. Using the full Spanish MEACorpus 2023 (5,129 samples), the model outperformed the baselines. The data are unambiguous: the fuzzy engine adds real value. It corrected 168 errors unresolved by linear fusion, 287 missed by the text module, and 162 cases where both baselines failed simultaneously. This shows that combining prosody and ASR confidence with linguistic rules helps recover correct predictions under noisy conditions.

Importantly, this phase establishes a clear heuristic benchmark. The pipeline achieves 43.8% recall and 0.429 weighted F1 with real Whisper transcriptions. This is not an upper bound—only the baseline that neural models must surpass under the same conditions. Notably, the simple heuristic audio classifier has a low F1 (0.213) by design, which helps demonstrate that the framework's auditing capabilities remain valid regardless of classifier strength.

Regarding Immutable Explainability, the system meets expectations: the fuzzy engine produces a readable XAI trace, and the `txid` anchored on Sepolia provides independent blockchain-based verification. This breaks the "data silo" problem: AI operation and auditability become decoupled in a decentralized registry. Validating the anchoring Proof of Concept establishes the foundation for future user-sovereign architectures (SSI).

7.2 Future Work. Four clear lines of work emerge:

(i) Neural Models and Empathetic Response: Replace heuristic classifiers with SOTA models (fine-tuning BETO/RoBERTa-es, Wav2Vec2/HuBERT) to surpass the heuristic benchmark. Compare results with EmoSPeech IberLEF 2024 baselines to position the architecture within the Spanish-language landscape (Mares et al., 2025).

(ii) Empathetic Response Generation: Incorporate a response generator (NLG), likely LLM-based (Zhang et al., 2025), conditioned on detected emotion (e.g., anger). This will enable contextual and affective dialogue, completing the interaction loop.

(iii) Expansion of the Sovereignty Layer (Blockchain/SSI):

After validating Sepolia anchoring as a decentralized root of trust, the next step is a full DID/VC workflow, allowing users to manage their encrypted `fusion_details` off-chain and interact with smart contracts that verify proofs on-chain without revealing sensitive data. This transitions the system from a "proof of existence" model to a "verifiable credentials" ecosystem.

(iv) Realistic Evaluation (Users, Models, Cross-Corpus):

Re-evaluate the pipeline with neural models under real-world conditions: deep classifiers for text and audio, quantifying the effect of Word Error Rate and extending evaluation to additional Spanish corpora.

All of this work forms part of the ongoing master's thesis (Fransoy, in preparation) and has already been presented in specialized workshops (Fransoy et al., 2025). The entire environment is fully reproducible (Docker, commit hash), and the source code is available at: https://github.com/fransoymarcelo/immutable-explainability-heuristic-model.

**Funding**
This research received no specific grant from any funding agency in the public, commercial, or not-for-profit sectors.

## Referencias

Bhardwaj, T., & Sumangali, K. (2025). An explainable federated blockchain framework with privacy-preserving AI optimization for securing healthcare data. Scientific Reports, 15, 21799. https://doi.org/10.1038/s41598-025-04083-4

Bohus, D., & Horvitz, E. (2019). Situated Interaction. In The Handbook of Multimodal-Multisensor Interfaces, Volume 3 (pp. 105–143). ACM and Morgan & Claypool.

Devlin, J., Chang, M.-W., Lee, K., & Toutanova, K. (2019). BERT: Pre-training of deep bidirectional transformers for language understanding. NAACL-HLT.

Faiury, A., et al. (2025). Enhancing speech emotion recognition with graph-based multimodal fusion and prosodic features. Proc. Interspeech 2025.

Fayek, H., Lech, M., & Cavedon, L. (2022). Evaluating Wav2Vec 2.0: A study on speech representation learning. IEEE/ACM Transactions on Audio, Speech, and Language Processing, 30, 1348–1359.

Feng, Y., Li, Y., & Wang, L. (2024). ASR-robust speech emotion recognition with multimodal fusion. ICASSP 2024. IEEE.

Fransoy, M. (en preparación). Interacción por voz afectiva y segura: un enfoque híbrido de PLN, Lógica Difusa y Blockchain. Tesis de Maestría, UTN FRBA.

Fransoy, M., Hossian, A., Merlino, H., & Pollo Cattaneo, M. F. (2025). Interacción por voz afectiva y segura: un enfoque híbrido de PLN, Lógica Difusa y Blockchain. WICC 2025.

Hao, Y., & Liu, D. (2025). Neuro-fuzzy systems for interpretable deep learning: A comprehensive survey. Neural Computing and Applications, 37(5), 8765–8798.

He, L., et al. (2024). MF-AED-AEC: Speech emotion recognition by leveraging multimodal fusion, ASR error detection, and ASR error correction. ICASSP 2024.

Hsu, W.-N., Bolte, B., Tsai, Y.-H. H., Lakhotia, K., Salakhutdinov, R., & Mohamed, A. (2021). HuBERT: Self-supervised speech representation learning by masked prediction of hidden units. IEEE/ACM Transactions on Audio, Speech, and Language Processing, 29, 3451–3460.

Jang, J.-S. R. (1993). ANFIS: Adaptive-network-based fuzzy inference system. IEEE Transactions on Systems, Man, and Cybernetics, 23(3), 665–685.

Kulothungan, V. (2025). Using blockchain ledgers to record AI decisions in IoT. IoT, 6(3), 37. https://doi.org/10.3390/iot6030037

Kushwaha, A. S. (2025). Blockchain-based logging for auditing AI decisions. Scientific Journal of Artificial Intelligence and Blockchain Technologies, 2(2), 10–19. https://doi.org/10.63345/sjaibt.v2.i2.302

Latif, S., et al. (2023). Privacy-preserving speech processing: A tutorial. arXiv preprint arXiv:2305.05227.

Lawrence, D., et al. (2024). Artificial intelligence in mental health: A review of the ethical challenges. The Lancet Digital Health, 6(1), e56–e65.

Liu, Y., Ott, M., Goyal, N., Du, J., Joshi, M., Chen, D., ... & Stoyanov, V. (2019). RoBERTa: A robustly optimized BERT pretraining approach. arXiv preprint arXiv:1907.11695.

Mares, A., et al. (2025). Advancing Spanish speech emotion recognition: A comprehensive benchmark of pre-trained models. Applied Sciences, 15(8), 4340.

OpenAI. (2022). Whisper: Robust speech recognition via large-scale weak supervision [Model]. https://github.com/openai/whisper

Pan, R., et al. (2024). Overview of EmoSPeech at IberLEF 2024: Multimodal speech-text emotion recognition in Spanish. Procesamiento del Lenguaje Natural, 73, 359–368.

Pan, R., et al. (2024). Spanish MEACorpus 2023: A multimodal speech–text corpus for emotion analysis in Spanish from natural environments. Expert Systems with Applications, 254, 124317.

Parlak, İ. E. (2025). Blockchain-assisted explainable decision traces (BAXDT): An approach for transparency and accountability in artificial intelligence systems. Knowledge-Based Systems, 329, 114402.

Pegwar, T., & Siddiqui, R. (2025). Blockchain + AI for transparent and auditable AI models. International Journal of Latest Technology in Engineering, Management & Applied Science, 14(13), 57–61.

Pei, Z., et al. (2024). Affective computing: Recent advances, challenges, and future trends. iComputing, 2(1), e0076.

Rahman, W., Hasan, M. K., Lee, S., Bagher Zadeh, A., Mao, C., Morency, L.-P., & Hoque, E. (2020). Integrating multimodal information in large pretrained transformers. Proceedings of the 58th Annual Meeting of the Association for Computational Linguistics,

2359–2369. https://doi.org/10.18653/v1/2020.acl-main.214

Sahu, S., et al. (2019). How does ASR performance affect speech emotion recognition? Interspeech 2019, 4365–4369.

Torres, A., & Nieto, J. J. (2006). Fuzzy logic in medicine and bioinformatics. Journal of Biomedicine and Biotechnology, 2006.

Van, T., et al. (2025). Multimodal emotion recognition in conversations: A survey. EMNLP 2025 Findings. arXiv preprint arXiv:2505.20511.

Zadeh, L. A. (1975). The concept of a linguistic variable and its application to approximate reasoning. Information Sciences, 8(3), 199–249.

Zao-Sanders, M. (2025, abril). How people are really using gen AI in 2025. Harvard Business Review. https://hbr.org/2025/04/how-people-are-really-using-gen-ai-in-2025

Zhang, J., et al. (2025). Affective computing in the era of large language models: A survey from the NLP perspective. arXiv preprint arXiv:2408.04638.

Zichichi, M., et al. (2024). A survey of blockchain-based privacy applications: An analysis of consent management and self-sovereign identity approaches. arXiv preprint arXiv:2411.16404.